\documentstyle[amsmath,amssymb,11pt,epsfig]{article}
\begin{document}
\title{\bf{Leptonic universality breaking in $\Upsilon$ decays as
a probe of new physics}
\thanks{Research under grant FPA2002-00612.}}
\author{
Miguel Angel Sanchis-Lozano$^{a,b}$\thanks{Email: Miguel.Angel.Sanchis@uv.es}
\vspace{0.4cm}\\
(a) Instituto de F\'{\i}sica
Corpuscular (IFIC), Centro Mixto Universidad de Valencia-CSIC \\
(b) Departamento de F\'{\i}sica Te\'orica, Universidad de Valencia \\
Dr. Moliner 50, E-46100 Burjassot, Valencia (Spain)}
\date{}
\maketitle
\begin{abstract}
	In this work we examine the possible existence of 
        new physics beyond the standard model which could  
        modify the branching fractions of  
        the leptonic (mainly tauonic) decays of 
        bottomonium vector resonances below the $B\bar{B}$ threshold. 
        The decay width is factorized as the product of two pieces: 
        a) the probability of an intermediate pseudoscalar color-singlet
        $b\bar{b}$ state (coupling to the dominant Fock state of 
        the Upsilon via a magnetic dipole transition) and a soft 
        (undetected) photon; b) the annihilation width of the $b\bar{b}$ 
        pair into two leptons, mediated by a non-standard CP-odd Higgs 
        boson of mass about $10$ GeV, introducing a quadratic
        dependence on the lepton mass in the partial width. The process 
        would be unwittingly ascribed to the $\Upsilon$ leptonic channel
        thereby (slightly) breaking lepton universality.
        A possible mixing of the pseudoscalar Higgs and bottomonium
        resonances is also considered. 
        Finally, several experimental signatures to check out the
        validity of the conjecture are discussed.
\end{abstract}
\vspace{-14.5cm}
\large{
\begin{flushright}
  IFIC/03-36\\
  FTUV-03-0725\\
  July 25, 2003\\
  hep-ph/0307313
\end{flushright} }
\vspace{14cm}
\begin{small}
PACS numbers: 14.80.Cp, 13.25.Gv, 14.80.-j \\
Keywords: Non-standard Higgs, New Physics, bottomonium leptonic decays, 
lepton universality
\end{small}
\newpage

\section{Introduction}

The Standard Model (SM) has become nowadays the necessary reference
to confront experimental data with theory: any possible discrepancy
between them  is commonly denoted as New Physics (NP), actually 
implying the need for
some new assumptions or extensions of the basic physical postulates.
In the SM, matter and gauge fields follow different statistics, the
former being fermions and the latter bosons. As is well-known,
there are important
reasons to believe that this is quite unsatisfactory. One
of the major motivations to extend the SM is to resolve the hierarchy
and fine-tuning questions between the electroweak scale and the
Planck scale. Supersymmetry is a very nice solution in this regard,
since diagrams with superpartners exactly cancel the quadratic 
divergences of the SM diagrams. In particular, it requires
that the boson sector of the SM including the
Higgs structure should be enlarged with new scalar fields. 
Until recently, supersymmetry was thought as the only
possibility to solve the hierarchy problem, partly because of the 
lack of known alternatives. However, a new formulation 
for the electroweak symmetry breaking (dubbed $\lq\lq$little
Higgs'' theories \cite{arkani01}) has recently emerged 
where cancellation of divergences 
occur, conversely to supersymmetry, 
between particles with the same statistics. The (initially massless)
Higgs fields can be seen in this framework 
as Goldstone bosons, adquiring a mass and becoming pseudo-Goldstone
bosons via explicit symmetry breaking at the electroweak scale, but still
protected by an approximate global symmetry which would keep them
relatively light.

This is not the whole story, however. Another approach to solve 
the hierarchy problem aimed to the old idea on Kaluza-Klein
extra dimensions, either in
an ADD scenario \cite{arkani,antoniadis} or in a 
Randall-Sundrum model \cite{randall}.
In both cases, the scalar sector would be increased by the
presence of neutral bosons
like scalar gravitons and radions (the latter associated to
the quantum oscillations of the interbrane separation), eventually 
leading to measurable deviations from the SM \cite{han,giudice}. 
Moreover, let us cite 
another example of this kind of extensions of the SM: the axion
model, originally introduced in \cite{wilczek,weinberg} as a 
consequence of the spontaneous breaking of the global $U(1)_{PQ}$ 
axial symmetry. Nowadays, the
axion is a pseudoscalar field
appearing in a variety of theories with different meanings including
superstring theory, yielding
sometimes a massless particle and others a massive one: in
astrophysics the axion represents a good candidate of
the cold dark matter component of the Universe.

Although there are well established mass bounds 
(e.g. from LEP searches \cite{lep})
for the standard Higgs boson and the Minimal Supersymmetric
Standard Model (MSSM), the situation can be
different in more general scenarios   
where the tight constraints on the parameters of the theory do not
apply, leaving still room for
light Higgs bosons compatible with 
present data although the point is currently 
controversial (see, for example, \cite{opal,maria1,maria2,cheung03}).
As a suggestive example, let us mention
that a possible mixing between Higgs bosons from a two doublet
structure (which will deserve special attention in this paper)
and resonances would
alter the exclusion limits put by LEP data on light
Higgs bosons since the branching fraction (BF) into tau pairs
decreases considerably in this case \cite{opal,drees}. 
Moreover, in the so-called next to MSSM
(NMSSM) a gauge singlet superfield is added to the
MSSM spectrum \cite{nilles,ellis}: new CP-even and CP-odd Higgs bosons
enter the game. For some choices of the parameters of this model, one can
obtain very light pseudoscalar Higgs states evading the LEP constraints and
whose detection might require some dedicated efforts at the LHC
\cite{hugonie}.         

The search for axions or light Higgs bosons in the decays of 
heavy resonances has several attractive features: 
first, the couplings of the former to fermions are 
proportional to their masses and therefore enhanced with respect to 
lighter mesons. Second, theoretical calculations are more reliable, 
notably with the advent of 
non-relativistic quantum chromodynamics (NRQCD) \cite{caswell,bodwin}.
Indeed, intensive searches for a light Higgs-like boson (to be generically
denoted by $\phi^0$ in this paper)  have been
performed according to the so-called Wilczek mechanism \cite{wilczek77}
in the radiative decay of vector heavy quarkonia like
the Upsilon resonance (i.e. $\Upsilon{\rightarrow}\ \gamma \phi^0$). 
To date,
none of all these searches has been successful, but have 
provided valuable constraints on the mass values of light Higgs 
bosons \cite{gunion}.

Nevertheless, in this paper we develop the key ideas already presented in
\cite{mas02,mas202} on a possible signal
of NP based on the $\lq\lq$apparent'' breaking of  
lepton universality in 
bottomonium decays. {\em Stricto sensu}, lepton universality
implies that the electroweak couplings 
to gauge fields of all charged 
lepton species should be the same; according to our
interpretation, the (would-be) dependence
on the leptonic mass of the leptonic branching fractions
(${\cal B}_{\ell\ell}$ ; $\ell=e,\mu,\tau$) 
of $\Upsilon$ resonances below the $B\bar{B}$ threshold, if 
experimentally confirmed by forthcoming measurements, might be viewed
as a hint of the existence of a quite light Higgs - of mass about 10 GeV -
deserving a closer look.

\subsection{Two-Higgs Doublet Models}

In its minimal version, the SM requires a complex scalar weak-isospin
doublet to spontaneously break the electroweak gauge symmetry. 
As already commented, theories
that try to resolve the hierarchy and fine-tuning problems
imply the extension of the Higgs sector.
Loosely speaking, the simplest way (i.e. adding the fewest number of arbitrary
parameters) corresponds to assume an extra Higgs doublet, i.e.
the Two-Higgs Doublet Model (2HDM) \cite{gunion}.
The Higgs content of this theory is the following: a charged pair ($H^{\pm}$),
two neutral CP-even scalars ($h^0$, $H^0$) and a neutral
CP-odd scalar ($A^0$) often referred as a pseudoscalar.
Let us also note that diverse extended
frameworks beyond the SM can lead to an effective theory
at low energies equivalent to the 2HDM. On the other hand, there exist 
models with higher representations for the Higgs
sector (e.g. Higgs triplets or the above-mentioned NMSSM) 
leading to more complicated structures \cite{gunion02}. 

Any two-doublet Higgs model has to cope with
the potential problem of enhancing the flavor-changing neutral
currents (FCNC). Several solutions have been proposed to overcome 
this serious difficulty. In 
the Type-I 2HDM only one
of the Higgs doublets couple to quarks and leptons and, 
since the process which diagonalizes the mass matrix of quarks
equally can diagonalize the Higgs coupling, there is
no flavor-changing vertex for the Higgs bosons at the end. 
(Note that in such case
the Higgs coupling to $b$ quarks is not enhanced.) 
Another extreme possibility to avoid FCNC's is based
on the assumption that one Higgs doublet does not
couple to fermions at all whereas the other Higgs' couples to
fermions in the same way as in the minimal Higgs model.
On the other hand, 
the Type-II 2HDM allows one of the Higgs doublet couple to the
up quarks and leptons while the other Higgs doublet can couple to 
down-type quarks and leptons.
This is the kind of model on which
we shall focus in the following, excluding MSSM
\footnote{The Higgs sector
of the MSSM can be viewed as a particular
realization of a constrained Type II 2HDM  with
less parameters free. However, in this paper we are not
considering the 2HDM as a low-energy approximation of
the MSSM, but in more general grounds.}
since current limits rule out a very light pseudoscalar Higgs 
boson \cite{lep} as advocated along this work. Nevertheless, other 
alternative scenarios
as those mentioned in the Introduction can not be discarded.

Among other new parameters of the 2HDM, one of special phenomenological
significance in this work
is the ratio of the vacuum expectation values ($v_{1,2}$ of the Higgs
down- and up-doublets respectively) usually
denoted as $\tan{\beta}=v_2/v_1$, where
$v_1^2+v_2^2=v^2$ with $v \simeq 246$ GeV fixed by the $W$ mass. 
Indeed, $\tan{\beta}$ governs the Yukawa couplings between 
Higgs bosons and fermions, thereby potentially enhancing the rate 
of processes forbidden by the SM, but
allowed thanks to new contributions heralding the existence of NP.

The layout of the paper is the following: in section 2 we tentatively
introduce the hypothesis of a light non-standard Higgs boson which could 
modify the leptonic decay rate
of $\Upsilon$ resonances; in section 3 we firstly apply 
time ordered second-order
perturbation theory for a two-step process: prior photon
radiation from the Upsilon leading to
a pseudoscalar intermediate state followed by
its annihilation into a lepton pair mediated
by a CP-odd Higgs. Alternatively, we consider in subsection 3.2 the  
factorization of the decay width assuming the existence 
of Fock states in hadrons containing
(ultra)soft photons as low-energy degrees of freedom in analogy to
gluons in NRQCD. In sections 4 and 5, we focus on a 2HDM(II) model
and the effects of the postulated NP
contribution on the leptonic branching fraction are analyzed
in the light of current experimental data: we 
conclude from a statistical test 
that lepton universality can be rejected 
at a 10$\%$ level of significance. Possible mixing
between a pseudoscalar Higgs and $\eta_b$ resonances is also
considered, and its consequences on
the hyperfine splitting between vector and pseudoscalar states. We 
finally gather technical details in three appendices at the end of the paper.

\section{Searching for a light Higgs-like boson in $\Upsilon$ leptonic decays}

The starting point of our considerations
is the well-known Van Royen-Weisskopf formula 
\cite{royen} including the color factor \footnote{As is well-known,
gluon exchange in the short range part of the quark-antiquark potential
makes significant corrections to Eq.(1) \cite{leader}, but
without relevant consequences in our later discussion
as we focus on relative differences between leptonic decay modes.} for  
the leptonic width of a vector quarkonium state 
without neglecting leptonic masses, 
\begin{equation}
{\Gamma}^{(em)}_{\Upsilon\to\ell\ell}\ =\ 
4\alpha^2Q_b^2\ \frac{|R_n(0)|^2}{m_{\Upsilon}^2}\ {\times}\ 
K(x_{\ell})
\label{eq:Gamma0}
\end{equation}
where $\alpha\ {\simeq}\ 1/137$ is the electromagnetic fine 
structure constant;
$m_{\Upsilon}$ denotes the mass of the vector particle 
(a $\Upsilon(nS)$ resonance in this particular case) and
$Q_b$ is the charge of the relevant (bottom) quark ($1/3$ in units of $e$);
$R_n(0)$ stands for the non-relativistic radial wave function of the
$b\overline{b}$ bound state at the origin; finally, the
$\lq\lq$kinematic'' factor $K$ reads
\begin{equation}
K(x_{\ell})\ =\ (1+2x_{\ell})(1-4x_{\ell})^{1/2}
\end{equation}
where $x_{\ell}=m_{\ell}^2/m_{\Upsilon}^2$.
Leptonic masses are usually neglected in Eq.(1) (by setting $K$
equal to unity) except for the decay into $\tau^+\tau^-$ 
pairs. Let us note that $K(x_{\ell})$ is a decreasing function of 
$x_{\ell}$:
the higher leptonic mass the smaller decay rate. 
Such $x_{\ell}$-dependence is quite weak for bottomonium and, consequently,
we will assume that lepton universality implies the constancy
of the width (1) for all lepton species.

However, in this work we are conjecturing the existence 
of a light Higgs-like 
particle whose mass would be close
to the $\Upsilon$ mass and which could show up in the cascade decay:
\begin{equation}
\Upsilon\ {\rightarrow}\ \gamma\ \phi^0({\rightarrow}\ \ell^+\ell^-)\ \ \ ;\ 
\ \ \ell=e,\mu,\tau
\end{equation}
Actually, this process may be seen
as a continuum radiative transition that in principle permits the
coupling of the bottom quark-antiquark pair to a particle 
of variable mass and $J^{PC}$:
$0^{++}, 0^{-+}, 1^{++}, 2^{++}...$ (always positive charge conjugation).
In this investigation we will confine our attention to the two 
first possibilities: a scalar or a pseudoscalar boson. In fact,  
intermediate bound states and not the continuum will play
a leading role in the process, as we shall see. In the language
of perturbation theory, a magnetic dipole transition (M1) 
would yield at leading order a pseudoscalar
$b\bar{b}$ state from the initial-state
vector resonance, subsequently annihilating into
a dilepton. Alternatively, there should be a certain probability
that a pseudoscalar color-singlet $b\bar{b}$ system could exist in 
the Fock decomposition of the physical Upsilon state, as the light
degrees of freedom would carry the remaining quantum numbers.

Throughout this work, we will focus on vector $\Upsilon$ states 
of the bottomonium family below open
flavor \footnote{The $\Upsilon(3S)$ state is excluded in the
present analysis since only experimental data for the muonic channel \cite{pdg}
are currently available; see http://pdg.lbl.gov for regular updates.},
and the complete process (3) actually would be
\begin{equation}
\Upsilon \rightarrow \gamma_s\,b\bar{b}[n](\rightarrow \phi^0\rightarrow
\ell^+\ell^-)\,\,\ ;\,\,\ \ell=e,\mu,\tau
\label{eq:channel}
\end{equation}
where  $\gamma_s$ denotes a 
{\em soft} (= unobserved) photon and
$b\bar{b}[n]$ stands for those intermediate states of
different quantum numbers collectively denoted by $n$, either 
on the continuum or as bound states.  
Note that $\phi^0$ is not a
real particle in the channel (\ref{eq:channel}), 
conversely to the Wilczek mechanism \cite{wilczek77}, 
but a virtual state mediating the annihilation of a 
$b\bar{b}$ intermediate state into the final-state lepton pair.
Hence, if the $\phi^0$ mass were quite close to 
the $\Upsilon$ mass, the Higgs propagator could enhance 
significantly the width of whole process. In fact, 
the analysis performed by OPAL \cite{opal}
using LEP data assuming a mixing between Higgs bosons and 
bottomonium resonances \cite{drees} does not permit to exclude a light Higgs 
of mass around 10 GeV using reasonable 
$\tan{\beta}$ values in the 2HDM(II). Moreover, 
one should not dismiss other possible 
scenarios, as pointed out in the Introduction.

Since the radiated photon would escape 
detection in our guess
\footnote{Experimental measurements of ${\cal B}_{\ell\ell}$
include soft radiated photons which, however, have to be taken into 
account for a consistent definition of the leptonic widths \cite{pdg}, 
as we claim for the NP contribution advocated in this work.},
the NP channel (\ref{eq:channel})
would be unwittingly ascribed to the leptonic decay mode of the
Upsilon resonance, introducing in its ${\cal B}_{\ell\ell}$
a quadratic leptonic mass dependence opposite
to that of Eq.(2) due to the Higgs coupling to fermions
This is a cornerstone in our conjecture
but likely of practical significance only for the $\tau^+\tau^-$
decay mode, where 
missing energy is experimentally required as one of the selection 
criteria\cite{cleo94}: events with
photons of order 100 MeV would be included in
the sample of tauonic decays, ultimately contributing to the measured 
leptonic BF. On the contrary, the electronic and muonic
BF's would be affected to a much lesser extent,
both because of: i) the smaller leptonic mass; 
ii) the experimental constraint on the 
reconstructed dilepton invariant mass, which
restricts severely the energy of possible $\lq\lq$lost'' photons
\footnote{The leptonic mass squared with a final-state
photon is given by 
$m_{\ell\ell}^2=m_{\Upsilon}^2(1-2E_{\gamma}/m_{\Upsilon})$.
Hence $E_{\gamma}$ is much more limited by invariant mass reconstruction
of either electrons or muons than for tau's where such 
constraint is not applicable. I especially acknowledge N. 
Horwitz and the CLEO collaboration for correspondence in this regard.}.

\section{Intermediate $b\bar{b}$ pseudoscalar states}
\label{sec:cascade}

Along this section, we examine the role played by
intermediate states in the process (4) 
according to two different schemes: firstly, we
use time-ordered perturbation theory (TOPT) to deal 
with the formation, via an electromagnetic
transition from the initial-state $\Upsilon$, of a virtual $b\bar{b}$ state 
and its subsequent annihilation into a lepton pair.
As an alternative approach, we rely on the separation between 
long- and short-distance physics following the main lines of 
a non-relativistic effective theory (NRET) like              
NRQCD \cite{bodwin} - albeit replacing a gluon 
by a photon in the usual Fock decomposition 
of hadronic bound states - and NRQED \cite{labelle,soto}.
Different results for the
final widths in each approach come up, however; they are
discussed in subsection 3.3. Finally, let us
note that, despite we generically refer to the Upsilon ($\Upsilon$)
in our study,
actually we are focusing on the $\Upsilon(1S)$ state 
because of more 
precise data on its tauonic BF (${\cal B}_{\tau\tau}$)
w.r.t. the $\Upsilon(2S)$, and not yet available
for the $\Upsilon(3S)$.

\subsection{Time-ordered perturbative calculation}

Let us write the amplitude for the process (\ref{eq:channel})
using TOPT at lowest order:
\begin{equation}
T_{\,\Upsilon\to\gamma_s\,\ell\ell}=\sum_{n}
\frac{\langle \, \ell^+\ell^- | \,{\mbox H}\, | \,n \,\rangle 
\langle \, \gamma_s\,n \,| \,{\mbox H}\, | \,\Upsilon\, \rangle}
{m_{\Upsilon}-E_n-k+i\epsilon}
\label{eq:TOPT}
\end{equation}
The sum extends over all possible $b\bar{b}$ intermediate states with 
proper quantum numbers and energy $E_n$,
and $k$ is the energy of the unobserved 
photon. Bottomonium states in a $^1S_0$ configuration should dominate
the sum, as we are facing the radiative decay of a $\Upsilon$ resonance. 
Intermediary states of higher angular momentum will be suppressed, since
the associated electromagnetic transitions would involve 
higher multipole moments; M1 
transitions to $S$-wave states with different
principal quantum numbers  
will neither be considered, as they involve orthogonal wave functions in
the non-relativistic limit.

From expression (\ref{eq:TOPT}) we also see that intermediate states 
with energies closer to
the $\Upsilon$ mass are enhanced with respect to those of higher
virtuality. In that sense, the continuum $b\bar{b}$ contribution,
starting with a pair of $B\bar{B}$ mesons ($E_{B\bar{B}}\simeq10.56$ GeV),
is well above the $\Upsilon(1S)$ and $\Upsilon(2S)$ masses, and the main
contributions to the decay would come from intermediate
pseudoscalar $S$-wave bound states, i.e. the $\eta_b$ 
resonances, differing from the $\Upsilon$ resonances in virtue of the
hyperfine structure.

Squaring the amplitude (\ref{eq:TOPT}) and including 
phase space integrations, the width of the process reads  
\begin{eqnarray}
\Gamma_{\,\Upsilon\to\gamma_s\,\ell\ell}&=& \int dk\,\rho_{\gamma}
\left|\frac{\langle \,\eta_b\, \gamma_s \,| \,{\mbox H}\, | \,\Upsilon\, 
\rangle}{m_{\Upsilon}-E_{\eta}-k+i\epsilon}\right|^2
\nonumber\\[2mm]
&&\times\int dE_{\ell\ell}\,\, \rho_{\ell^+\ell^-}
\,|\langle \, \ell^+\ell^- | \,{\mbox H}\, | \,\eta_b \,\rangle |^2\,
2\pi\,\delta(m_{\Upsilon}-k-E_{\ell\ell})+\dots
\label{eq:width1}
\end{eqnarray}
where the dots stand for other intermediate states contributing to the sum in 
(\ref{eq:TOPT}).  For explicit expressions of the particle densities
$\rho_{\gamma}$, $\rho_{\ell^+\ell^-}$, we refer the reader to \cite{sakurai}.
The last integral amounts to the width of a $0^{-+}$ resonance
with mass $m_{\eta_b^*}=m_{\Upsilon}-k$ decaying into a pair of leptons, 
which we will
denote as $\Gamma_{\eta_b^*\to\ell\ell}$. Section \ref{sec:higgs}
will be devoted to the calculation of this annihilation via the
proposed Higgs exchange.

The matrix element squared for the M1 transition between 
spin-triplet and spin-singlet  
$S$-wave states of quarkonium
can be written in terms of the $\Upsilon(nS)\to\gamma_s\, \eta_b^*(nS)$ width
after performing the integration over allowed photon states
\footnote{Notice that there is no
infrared singularity in the case of a 
{\em magnetic} dipole radiation.} with energy $k$,
\begin{equation}
2\pi\int \rho_{\gamma}\,
\left|\langle \,\eta_b\, \gamma_s \,| \,{\mbox H}\, | \,\Upsilon\,
\rangle\right|^2\ 
=\ \Gamma^{\,M1}_{\,\Upsilon\to\gamma_s\,\eta_b^*}\ 
\simeq\ \frac{16\alpha}{3}\Big( \frac{Q_b}{2m_b} \Big)^2k^3
\end{equation}
where we emphasize the off-shellness of pseudoscalar resonance
by writing $\eta_b^*$.
In the last step, we used a non-relativistic approximation to estimate the
width $\Gamma^{\,M1}_{\,\Upsilon\to\gamma_s\,\eta_b^*}$ \cite{oliver}. We 
shall return later to this formula in greater detail. 
Plugging the above result into Eq.(6) one obtains 
\begin{equation}
\Gamma_{\,\Upsilon\to\gamma_s\,\ell\ell}\ =\ 
\frac{16\alpha}{3}\Big( \frac{Q_b}{2m_b} \Big)^2\frac{1}{2\pi}
\int_0^{\Lambda} dk\,\frac{k^3}{\left| m_{\Upsilon}-E_{\eta_b}-k+i\epsilon\right|^2}\,
\,\Gamma_{\eta_b^*\to\ell\ell}
\label{eq:width2}
\end{equation} 
Under the {\it soft} photon hypothesis, we have
restricted the integration over photon energies to an upper limit 
$\Lambda\ll m_{\eta_b}$, 
dictated by both experimental photon energy resolution and event selection
criteria as pointed out in section 2.   
In that case we can safely 
retain the first term in the expansion of the $\eta_b$ energy, 
$E_{\eta_b}\ =\ \sqrt{m_{\eta_b}+k^2} \simeq m_{\eta_b}+k^2/2m_{\eta_b}+...$
in the denominator of (\ref{eq:width2}):
\begin{equation}
\Gamma_{\,\Upsilon\to\gamma_s\,\ell\ell}=
\frac{16\alpha}{3}\Big( \frac{Q_b}{2m_b} \Big)^2 \,
\bigg\{ \frac{1}{2\pi}
\int_0^{\Lambda} dk \,\frac{k^3}
{\left| m_{\Upsilon}-m_{\eta_b}+i\Gamma_{\eta_b}/2-k\right|^{\,2}}\
\Gamma_{\eta_b^*\to\ell\ell}\bigg\}
\label{eq:width3} 
\end{equation} 
The instability of the $\eta_b$ intermediate state has been taken into account
by the substitution $m_{\eta_b}\to m_{\eta_b}-i\Gamma_{\eta_b}/2$, valid in the
narrow-width case. A rough estimate of this width can be obtained
through the pQCD relation \cite{oliver}
$$
\frac{\Gamma_{\eta_b}}{\Gamma_{\eta_c}} \simeq \frac{m_b}{m_c}
\left[ \frac{\alpha_s(m_b)}{\alpha_s(m_c)} \right]^5
$$
With reasonable values of the quark masses, the running strong coupling
and the measured $\Gamma_{\eta_c}=16\pm 3\,$MeV \cite{pdg}, we get
$\Gamma_{\eta_b}\simeq 4\,$MeV.

Consider now the quantity between curly braces in Eq.(\ref{eq:width3}):
\begin{equation} 
\frac{1}{2\pi}
\int_0^{\Lambda} dk \,\frac{k^3}
{(m_{\Upsilon}-m_{\eta_b}-k)^2+\Gamma_{\eta_b}^2/4}\ 
\Gamma_{\eta_b^*\to\ell\ell}
\equiv {\cal I}(\Lambda,\Delta E_{hs})
\label{eq:PM1}
\end{equation}
The mass difference $m_{\Upsilon}-m_{\eta_b}\equiv \Delta E_{hs}$ is 
the hyperfine splitting between 
the $\Upsilon(nS)-\eta_b(nS)$ partners. Different approaches 
suggest that $\Delta E_{hs}$ lies in the interval $\sim 35-150$ MeV 
for $n=1$ (see \cite{aleph02} for a compilation of theoretical results). 
If  $\Lambda >
\Delta E_{hs}+\Gamma_{\eta_b}$, i.e. the range of energies for
the unobserved photon is higher that the mass difference $\Delta E_{hs}$, 
the integration region in Eq.(\ref{eq:PM1}) 
comprises the full $\eta_b$ resonance contribution; taking 
$\Lambda\simeq  \Delta E_{hs}$ would reduce the result by a half, and values
of $\Lambda$ below $\Delta E_{hs}-\Gamma_{\eta_b}$ would make 
${\cal I}$ almost vanishing.  Finally, a very large value of $\Lambda$ would
make (10) quadratically divergent (and then $\Lambda$ can be seen as an
ultraviolet regulator).

Now, keeping track of our discussion at the end of section 2, those
events with photons of energies up to several hundred MeV 
would pass the selection
criteria employed in experiments measuring the tauonic 
BF of $\Upsilon$ resonances. Therefore, one can safely 
take in this case ($\ell=\tau$) the upper limit $\Lambda$
in the integrals of Eqs. (\ref{eq:width2}-\ref{eq:PM1}) 
of the order of few hundreds of MeV, well above the 
quantity $\Delta E_{hs}+\Gamma_{\eta_b}$, but always $\Lambda \ll m_{\eta_b}$. 

We thus evaluate the integral in Eq.(\ref{eq:PM1}) and take the limit of  
small $\Gamma_{\eta_b}$, yielding
$$
{\cal I}(\Lambda,\Delta E_{hs})\ \simeq\ 
\frac{(\Delta E_{hs})^3}{\Gamma_{\eta_b}}\
\Gamma_{\eta_b\to\ell\ell}
$$
The off-shell width $\Gamma_{\eta_b^*\to\ell\ell}$ has been transformed
in this limit to the on-shell width $\Gamma_{\eta_b\to\ell\ell}$, as
$m_{\eta_b^*}=m_{\Upsilon}-\Delta E_{hs}=m_{\eta_b}$. 
In section 4 we will obtain an expression for this partial width
dominated by the postulated Higgs boson.

Finally, under the aforementioned assumption that the
photon energy cutoff $\Lambda$ 
is large enough, 
the partial width of the whole decay reduces to the factorized formula
\begin{equation}
\Gamma_{\,\Upsilon\to\gamma_s\,\ell\ell}\ =\ 
\Gamma^{\,M1}_{\,\Upsilon\to\gamma_s\eta_b}
\frac{\Gamma_{\eta_b\to\ell\ell}}{\Gamma_{\eta_b}}\ 
\simeq\ 
\frac{16\alpha}{3}\Big( \frac{Q_b}{2m_b} \Big)^2 \Delta E_{hs}^3\,
\frac{\Gamma_{\eta_b\to\ell\ell}}{\Gamma_{\eta_b}}
\label{eq:finalwidth} 
\end{equation} 
where $\Gamma^{\,M1}_{\,\Upsilon\to\gamma_s\eta_b}$ is now 
the on-shell M1 transition between real $\Upsilon(n)$ and
$\eta_b(n)$ states (i.e. $k=\Delta E_{hs}$). Dividing
both sides of Eq.(\ref{eq:finalwidth}) by 
the Upsilon total width $\Gamma_{\Upsilon}$,
one gets the final BF as the product of two BF's, namely
${\cal B}_{\,\Upsilon\to\gamma_s\,\ell\ell}= 
{\cal B}_{\Upsilon\to\gamma_s{\eta}_b} \times 
{\cal B}_{\eta_b\to{\ell\ell}}$.
We have thus proved that our approximation for the full process 
matches the corresponding
expression to a cascade decay taking place through a 
$\eta_b^*$ intermediate state above threshold. Remarkably, no dependence 
on the $\Lambda$ parameter is left, provided that the condition
$\Delta E_{hs}+\Gamma_{\eta_b} < \Lambda \ll m_{\eta_b}$  is satisfied.

As argued before, 
other possible intermediate contributions
(i.e. $B\bar{B}$ continuum) are numerically less relevant due to
their increasing mass difference
($m_{\Upsilon}-m_{B\bar{B}}$) in the denominator of Eq.(\ref{eq:TOPT}); 
in fact, one
can check that the interference between $\eta_b$ and continuum contributions
in Eq.(\ref{eq:width1}), assuming
that the matrix elements 
$\langle \,\eta_b\, \gamma_s \,| \,{\mbox H}\, | \,\Upsilon\,\rangle$, 
$\langle \,B\bar{B}\, \gamma_s \,| \,{\mbox H}\, | \,\Upsilon\,\rangle$
have the same $k$-dependence, gives a correction which is well below 1\%
for both the $\Upsilon(1S)$ and $\Upsilon(2S)$ resonances using
$\Delta E_{hs}\sim 100$ MeV and $\Lambda\sim 120$ MeV as reference
values. Pure continuum contributions would be even 
more suppressed.

\begin{figure}[htb]
\centerline{\hbox{
 \psfig{file=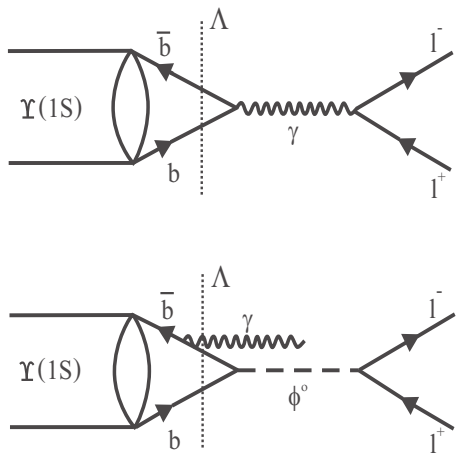,height=6.cm,width=6cm}  
}}
\caption{(a)[upper panel]: Electromagnetic 
annihilation of a $\Upsilon(1S)$ resonance into a charged
lepton pair through a virtual photon;
(b)[lower panel]: Hypothetical annihilation of a  
$b\bar{b}{[{}^1\!S_0^{(1)}]}$ state (existing either as a
Fock component of the $\Upsilon$ resonance,
or as a consequence of a M1 radiative transition) 
into a charged lepton pair through a Higgs-like particle 
(denoted by $\phi^0$).
The vertical dotted line represents the separation between long-distance
physics (on the l.h.s.) and short-distance physics (on the r.h.s.) 
corresponding to an arbitrary scale $\Lambda$ which
can be taken of the order of one hundred MeV, i.e. the 
soft photon energy.}
\end{figure}

\subsection{Long- and short-distance factorization according to NRET} 

Instead of supposing a final-state photon radiated by
the Upsilon via a M1 transition as in the precedent section, 
we will now consider the soft $\gamma_s$ incorporated as a 
dynamical photon into a Fock state of the resonance, in analogy to 
soft gluons in the framework of NRQCD. Admittedly, the
strong interaction rules the hadronic dynamics but  
electromagnetism still has
small but observable consequences, e.g. isospin breaking effects
\cite{ecker}.

In a naive quark model, quarkonium is treated as a nonrelativistic bound
state of a quark-antiquark pair in a static color field which sets up  
an instantaneous confining potential. Although this picture has been 
remarkably successful in accounting for the properties and
phenomenology of heavy quarkonia, it overlooks gluons whose wavelengths are
larger that the bound state size: dynamical gluons permit
a meaningful Fock decomposition (in the Coulomb gauge) of 
physical states
beyond the leading non-relativistic description with important
consequences both in decays and production of heavy quarkonium.
 
In particular, the $b\bar{b}$ system in a $\Upsilon$ vector
resonance can exist in a configuration other than the dominant 
$J^P=1^-$ since the 
light degrees of freedom mainly formed by gluons and
light quark-antiquark pairs
can carry the remaining quantum numbers, albeit with a
smaller probability. Moreover, it is arguable that
soft photons should be included among those low-energy hadronic 
modes too, therefore allowing the heavy quark-antiquark
system to be, for example, in a color-singlet, spin-singlet 
configuration, i.e. a $|b\bar{b}[{}^1\!S_0^{(1)}]+\gamma_s\rangle$ state
\footnote{We borrow this spectroscopic notation from NRQCD. 
Sometimes, the $b\bar{b}[{}^1\!S_0^{(1)}]$ state
will be denoted as $\eta_b^*$.}. Hence, such
dynamical photons 
would participate in some decay modes of heavy quarkonium, 
in analogy to dynamical gluons. 

According to NRQCD, higher Fock states containing soft gluons indeed
can participate actively in the decays of 
heavy quarkonium, as for example, its
annihilation into light hadrons \cite{bodwin,braaten98}. On the 
one hand, the probabilities of possible Fock components
are encoded in long-distance matrix elements; the 
heavy quark-antiquark annihilation itself, being a 
short-distance process, would be perturbatively calculable.
On the other hand, details about the complicated 
nonperturbative hadronization of gluons into final-state
light hadrons can be avoided in inclusive channels
by assuming the hadronization
probability equal to one.

Likewise, one can consider Fock states containing dynamical
photons, i.e. pertaining to the hadron during a long-time
scale as compared with the short-time process represented 
in our case by the $b\bar{b}$ annihilation
into a dilepton, of order $1/m_b$. Thus, a dynamical 
(quasi-real) photon, stemming 
from a $|b\bar{b}[{}^1\!S_0^{(1)}]+\gamma_s\rangle$ Fock state,
could end up (without need of hadronization)
as a final-state photon - though undetected
as postulated in this work. Obviously, important
differences stand between 
dynamical gluons and photons so that  
it will be instructive to review several
aspects of NRQCD relevant for our later development.

NRQCD is a low-energy effective theory for the strong interaction
removing the unwanted degrees of freedom associated to 
the heavy quark mass.  The following hierarchy between 
hadronic scales is usually assumed
in heavy quarkonium: $m_Q \gg m_Q\bar{v} \gg m_Q\bar{v}^2 \simeq\ 
{\Lambda}_{QCD}$, 
where $m_Q$ and $\bar{v}$ denotes the mass and 
relative velocity of the heavy quark 
\footnote{We employ the symbol $\bar{v}$ to denote the relative 
three-velocity instead of $v$, usual in NRQCD,
to avoid confusion with the vacuum expectation
value of the standard Higgs boson.}
respectively, and
$\Lambda_{QCD}$ is the strong interaction scale.
The bound state dynamics is chiefly
dominated by the exchange of Coulombic gluons with four-momentum
$(E \simeq m_Q\bar{v}^2,\vec{p} \simeq m\bar{v})$; soft gluons
have four-momenta of order $(m_Q\bar{v},m_Q\bar{v})$ and ultrasoft
gluons of order $(m_Q\bar{v}^2,m_Q\bar{v}^2)$. Likely the
above hierarchy makes sense for bottomonium since
$\bar{v}^2 \simeq 10^{-1}$; then the energy of ultrasoft
degrees of freedom turns out to be of order $500$ MeV or less.
In fact, the separation between 
energy scales is decisive in phenomenological applications of
NRQCD. 

Bodwin, Braaten and Lepage \cite{bodwin}
showed in a rigorous way that inclusive annihilation decays of 
heavy quarkonium can be 
factorized according to soft and hard
processes: a) Long-distance physics is encoded in 
matrix elements, providing
the probability for finding the heavy quark and antiquark in
a certain configuration within the meson which is suitable
for annihilation in each particular case; 
b) The short-distance annihilation of
the $Q\overline{Q}$ pair with given quantum numbers (color, spin,
angular momentum and total angular momentum) which could be
perturbatively calculated.
Therefore, the decay width is written as 
\begin{equation}
\Gamma(H)\ =\ 
\sum_n \tilde{\Gamma}_{Q\overline{Q}[n]}(\Lambda)\ 
\langle H|{\cal O}_n^H(\Lambda)|H \rangle
\label{eq:factor}
\end{equation}
where $\Lambda$ is an ultraviolet cutoff of the effective theory,  
separating the high- and low-energy scales. The 
short-distance coefficient
$\tilde{\Gamma}_{Q\overline{Q}[n]}(\Lambda)$ can be calculated
as a perturbation series in $\alpha_s(2m_Q)$.
The long-distance parameter $\langle H|{\cal O}_n^H|H \rangle$
determines the probability for the quarkonium to be in the 
$[n]$-configuration of the Fock decomposition
and can be interpreted as
an overlap between the $Q\overline{Q}[n]$ state and the final
hadronic state, requiring either a nonperturbative 
calculation (e.g. on the lattice) or
the extraction from experimental data. Similar arguments
apply to heavy quarkonium inclusive production 
(see, for example, \cite{mas01} and references therein).

In close analogy to the above procedure, let us also
introduce an energy parameter $\Lambda$ to separate short- 
from long-distance physics in the process under study in this work
as shown in Fig.1. (One might identify numerically
this $\Lambda$ parameter with
the upper limit of the integrals (8-10) in the precedent
section.) Thus, a 
factorization of the decay width of the $\Upsilon$ into two pieces
is applied: a) the probability ${\cal P}^{\Upsilon}(\eta_b^*\gamma_s)$ of 
the existence within the Upsilon of 
a $|b\bar{b}[{}^1\!S_0^{(1)}]+\gamma_s\rangle$ Fock 
state; b) the annihilation width 
${\Gamma}_{\eta_b^*\to\ell\ell}$ into a 
lepton pair via Higgs boson exchange. Therefore, 
$\Gamma_{\,\Upsilon\to\gamma_s\,\ell\ell}$
can be written in this approach as the product
\begin{equation}
\Gamma_{\,\Upsilon\to\gamma_s\,\ell\ell} =
{\cal P}^{\Upsilon}(\eta_b^*\gamma_s) 
\times {\Gamma}_{\eta_b^*\to\ell\ell}
\label{eq:factorization}
\end{equation}
This equation follows the spirit of the factorization 
given in Eq.(\ref{eq:factor}).
One important  difference, however,  
is that the long-distance quantity  
${\cal P}^{\Upsilon}(\eta_b^*\gamma_s)$ can be calculated
perturbatively in QED using a quark potential model 
(for the initial and
final state wave functions are involved) because of the smallness of 
the electromagnetic coupling $\alpha$, in contrast to 
NRQCD matrix elements.
On the other hand, the short-distance parameter 
${\Gamma}_{\eta_b^*\to\ell\ell}$ in Eq.(13)
 which we may identify with 
a particular $\tilde{\Gamma}_{Q\overline{Q}[n]}$ coefficient
in formula (12), can be calculated with the aid of the
Feynman rules of the model under consideration, as we shall later see.

\subsubsection{Estimate of the probability 
${\cal P}^{\Upsilon}(\eta_b^*\gamma_s)$}

In NRQCD the probabilities for
different Fock configurations of the heavy quark-antiquark pair 
in heavy quarkonium can be estimated
according to the number and order of the chromoelectric and chromomagnetic
transitions induced by the interaction effective Lagrangian,
needed to reach such states from the lowest configuration
or viceversa. Let us point out that in our particular case 
we are considering a mixed situation where
{\em electromagnetic} transitions occur between 
bound states of quarks which are mainly
governed by the strong interaction dynamics.
Moreover, another caveat is in order: we will compute
the transition rate between on-shell states, whereas the pseudoscalar
$\eta_b^*$ state in (13) should be somewhat off-shell. 
Nevertheless, we will assume this off-shellness small on account of
the low energy of the radiated photon. 

As commented at the beginning of section 3, we are focusing
on the $\Upsilon(1S)$ resonance, mainly because of
much more precise experimental data on its tauonic BF as 
compared to the $\Upsilon(2S)$, 
as displayed in Table 1, and still
missing for the $\Upsilon(3S)$. In addition, 
the larger number of possible intermediate pseudoscalar 
bound states for the two latter resonances would make more complicated
the theoretical analysis as compared with the $\Upsilon(1S)$ state,
where only the lowest $b\bar{b}[{}^1\!S_0^{(1)}]$
Fock configuration (i.e. a single $\eta_b(1S)$ state) should
contribute to the final annihilation into a dilepton.
Therefore, a
textbook expression \cite{oliver} has been employed to calculate
the width corresponding to a transition between S-wave states,
i.e. a direct M1-transition between
the $\Upsilon(1S)$ and the $\eta_b(1S)$ resonances.  

The probability of the Fock state $|\eta_b^*\gamma_s\rangle$ 
$\lq\lq$inside'' a $\Upsilon(1S)$ resonance
is then estimated as the ratio,
\begin{equation}
{\cal P}^{\Upsilon}(\eta_b^*\gamma_s)\ \approx\ 
\frac{\Gamma^{M1}_{\Upsilon{\rightarrow}\gamma_s\eta_b}}
{\Gamma_{\Upsilon}}\ {\simeq}\ 
\frac{1}{\Gamma_{\Upsilon}}\ \frac{4\alpha Q_b^2}{3m_b^2}\ 
\Delta E_{hs}^3\ \times\ |{\cal M}_{\Upsilon : \eta_b}|^2\ \sim 10^{-4}
\label{eq:probability}
\end{equation}
using $m_b=M_\Upsilon/2 \simeq 5$ GeV,
and the hyperfine mass splitting between the $\Upsilon$
and $\eta_b$ states $\Delta E_{hs} \simeq 50$ MeV as
a reference value \footnote{In the language of a low-energy 
effective theory, those low-energy photons would be properly designed as 
ultrasoft in accordance with the energy scale hierarchy for 
bottomonium. The incorporation of quite
higher energy photons to the Fock decomposition 
of the resonance would appear 
more problematic as the typical lifetime of such Fock
states becomes comparable to the short time-scale of the annihilation
process.}.
The matrix element ${\cal M}_{\Upsilon : \eta_b}$ is defined as
\[
{\cal M}_{\Upsilon : \eta_b} = 
\int_0^{\infty}dr\ u_{\Upsilon}(r)\ j_0(kr/2)\ u_{\eta_b}(r)
\]
where $u_{i,f}(r)$ represents the reduced radial wave function of the
initial and final resonance respectively, 
and $j_0$ is the spherical Bessel function.
We have made in (\ref{eq:probability}) the reasonable
approximation: ${\cal M}_{\Upsilon : \eta_b} \approx 1$.
Let us remark, however, that this parameter
involves the wave functions of the $\Upsilon$ and $\eta_b$
resonances, actually constituting a nonperturbative
matrix element appearing in the long-distance part of the
factorized width, in analogy to conventional NRQCD. 

In the absence of current experimental data, it
is worth noting that 
the order-of-magnitude $\sim 10^{-4}$ of (\ref{eq:probability}) 
agrees with more elaborated calculations. For example, 
L\"{a}hde \cite{lahde} obtains for the partial width of
the process $\Upsilon(1S)\rightarrow\eta_b(1S)\gamma$ the value 7.7 eV
for $\Delta E_{hs} = 59$ MeV, 
which corresponds to a BF of $1.5 \times 10^{-4}$ in
accordance with (14).

Let us now 
compare the partial widths of the $\Upsilon(1S)$ decays
into three gluons and two gluons plus a photon 
by means of the ratio \cite{leader},
\begin{equation}
r = \frac{\Gamma(\Upsilon(1S) \rightarrow \gamma gg)}
{\Gamma(\Upsilon(1S) \rightarrow ggg)}\ \simeq\ \frac{4}{5}\frac{\alpha}
{\alpha_s}\biggl(1-2.6\frac{\alpha_s}{\pi}\biggr)\ \sim\ 10^{-2}
\label{eq:leader}
\end{equation}
at the energy scale $\mu^2=m_{\Upsilon}^2$. Actually, 
$r$ can be crudely viewed as the relative probability 
(or suppression factor)
of a $|b\bar{b}[{}^1\!S_0^{(1)}]+\gamma_s\rangle$ Fock state
w.r.t. the corresponding
$|b\bar{b}[{}^1\!S_0^{(8)}]+g\rangle$ Fock state in the
$\Upsilon$ resonance.
On the other hand, when the gluon energy is of order $m_b\bar{v}^2$
or less, the probability for the latter 
colored Fock state scales as
$\bar{v}^4$ according to the NRQCD scaling-velocity rules 
\cite{schuler,bodwin}. 
Thus, numerically
${\cal P}^{\Upsilon}(\eta_b^*\gamma_s) 
 \simeq  r \times \bar{v}^4 \sim 10^{-4}$, 
since $\bar{v}^2 \simeq 10^{-1}$ for bottomonium, showing the
same order-of-magnitude as Eq.(\ref{eq:probability}). 
Hence, we will assume that Eq.(14) provides a reasonable estimate of 
the probability for the existence of a $\eta_b^*$ state
within the Upsilon 
\footnote{One might also consider in our conjecture a 
$|b\bar{b}[{}^1\!S_0^{(1)}]+3g\rangle$ Fock state yielding
an unobserved hadronic system in the final state ($2\pi, \rho,...$). 
However, this 
contribution would be very much suppressed because of
the large virtuality of the intermediate $\eta_b^*$ state since certainly
$\Delta E_{hs}$ should be quite smaller than
the hadronic invariant mass.} as a function of $\Delta E_{hs}$. 
 
Combining the $\lq\lq$master'' formula (13) and equation (14),
the NRET approach leads to
\begin{equation}
\Gamma_{\,\Upsilon\to\gamma_s\,\ell\ell}\ =\ 
\frac{\Gamma^{\,M1}_{\,\Upsilon\to\gamma_s\eta_b}} 
{\Gamma_{\Upsilon}}\ \Gamma_{\eta_b\to\ell\ell}\
\simeq\ \frac{1}{\Gamma_{\Upsilon}}
\frac{16\alpha}{3}\Big( \frac{Q_b}{2m_b} \Big)^2 
\Delta E_{hs}^3\,\Gamma_{\eta_b\to\ell\ell}
\label{eq:finalwidth2} 
\end{equation} 
for the final decay width, to be compared 
with Eq.(\ref{eq:finalwidth}) obtained from TOPT.

\subsection{Discussion}

From Eqs.(\ref{eq:finalwidth}) and (\ref{eq:finalwidth2}) it becomes
apparent that the calculations of the final width 
$\Gamma_{\,\Upsilon\to\gamma_s\,\ell\ell}$ based upon 
TOPT or upon factorization {\em \`a la} NRET 
lead to different results. Indeed, 
the total width $\Gamma_{\eta_b}$ of the $\eta_b$ resonance 
appears in the
denominator of formula (\ref{eq:finalwidth}) after the approximations;
instead, the total width $\Gamma_{\Upsilon}$ of the $\Upsilon$ resonance 
appears in formula (\ref{eq:finalwidth2}). Since one expects
$\Gamma_{\Upsilon} \ll \Gamma_{\eta_b}$ \footnote{
Because the hadronic decay of a pseudoscalar resonance
via the annihilation of the $Q\bar{Q}$ pair
into two gluons should proceed at a higher rate than the corresponding decay
channel of the vector resonance via three gluons.}, 
the decay rate for the whole channel $\Upsilon\to\gamma_s\,\ell\ell$ 
stemming from the NRET factorization turns out to be
much larger than the decay rate obtained from TOPT. In other words, 
both approaches are not dual each other.

Actually, the fact that one gets different expressions
and hence dissimilar numerical values for the width
$\Gamma_{\,\Upsilon\to\gamma_s\,\ell\ell}$ in both methods
is not contradictory in itself for they are based on
distinct physical assumptions. The time-ordered  perturbative scheme
with $\eta_b^*$ production above threshold essentially implies
a cascade decay, i.e. photon emission and subsequent
annihilation of the intermediate hadronic state. Only
under this hypothesis and the narrow width approximation, 
the factorization given by Eq.(\ref{eq:finalwidth}) is justified. 
On the other hand,
the factorization {\em \`a la} NRET postulated in
Eq.(\ref{eq:finalwidth2}) assumes that
a pseudoscalar color-singlet $b\bar{b}$ state exists inside 
the $\Upsilon$ as a Fock state. Both starting points are 
different and the results stemming
from each framework need not coincide
\footnote{Let us make a pedagogical analogy with radioactive
nuclides: the factorization given in expression (\ref{eq:finalwidth}) 
amounts to the product of branching fractions in a cascade
decay; conversely formula (\ref{eq:finalwidth2}) 
corresponds to the
coexistence of a radioactive nuclide in some proportion 
with a stable isotope in nature. The decay
rate of a sample of this element
would be given by the fraction of the radioactive
isotope (i.e. the probability to find a radioactive atom in the sample,
analogous to ${\cal P}^{\Upsilon}(\eta_b^*\gamma_s)$)
multiplied by its decay rate.}. 

In fact, equivalent situations can be found, e.g. in inclusive
hadroproduction of heavy quarkonium at large transverse 
momentum \cite{nason}, when comparing
the color-singlet fragmentation mechanism
(where all three perturbative final-state gluons
are attached to the hard interaction Feynman diagram)
versus the color-octet mechanism (where the emission of two soft
gluons from a nonperturbative colored state
takes place on a long-time scale). Another example where 
higher Fock states can compete with perturbative calculations 
is the explanation given by Braaten and Chen 
of the long-standing $\rho-\pi$ puzzle of $J/\psi$ and
$\psi'$ decays \cite{braaten}. In general, their decays into light hadrons
should occur by the annihilation of the $c\bar{c}$ pair
into three gluons but the discrepancy with experiment is about
two orders of magnitude in this channel. Instead, they argue that the process
is dominated by the higher color-octet  
contribution $c\bar{c}[{}^3\!S_1^{(8)}]$, 
thereby annihilating into a light quark pair via
$c\bar{c} \to g^* \to q\bar{q}$. The suppression of
this decay mode for the $\psi'$ is attributed to a dynamical effect
which cancels the $c\bar{c}$ wavefunction
at the origin. (Alternatively, Brodsky and Karliner \cite{brodsky}
suggested that the
decay into $\rho\pi$ should proceed through the intrinsic 
charm components of the light mesons.)

Hereafter,  we will adopt the NRET factorization as given 
by the master equation (\ref{eq:factorization}) and the numerical estimates 
will be based on Eq.(\ref{eq:finalwidth2}). The low-energy
regime of the (ultra)soft photons provides confidence in this
approach \footnote{On the other hand, in using Eq.(\ref{eq:finalwidth})
instead of Eq.(\ref{eq:finalwidth2})
large values of $\tan{\beta}$ are required at the end of the calculation
to account for the leptonic BF rise with the lepton mass, 
leading to a $\eta_b$ width exceendingly large in contradiction with
the narrow width approximation made along the way 
to get (\ref{eq:finalwidth}). On the contrary, no inconsistencies 
of that kind arise when using the NRET approach to
interpret the conjecture made in this work.}.

\section{Effects of a light neutral Higgs on the leptonic 
decay width} 
\label{sec:higgs}

Following a general scheme, fermions are supposed to couple to the
$\phi^0$ Higgs field
according to a Yukawa interaction term in the effective Lagrangian,
\begin{equation}
{\cal L}_{int}^{\bar{f}f}\ =\ -\xi_f^{\phi}\ \frac{\phi^0}{v}
m_f\bar{f}(i\gamma_5)f 
\label{eq:A0coupling}
\end{equation}
where $\xi_f^{\phi}$ denotes a factor
depending on the type of the Higgs boson
and the specific theory
under consideration, which
could enhance the coupling with a fermion 
(quark or lepton) of type $\lq\lq$$f$'' and therefore plays a
crucial role in our conjecture. In particular, $\phi^0$ 
couples to
the final-state leptons proportionally to their masses, 
ultimately required because
of spin-flip in the interaction of a fermion with a 
(pseudo)scalar, thereby providing an experimental 
signature for checking the existence of a light Higgs in
our study.
Lastly, note that the
$i\gamma_5$ matrix stands only in the case of 
a pseudoscalar $\phi^0$ field.

In this paper we are tentatively assuming that the mass of the 
light Higgs sought stands 
close to the $\Upsilon$ resonances below $B\bar{B}$ production:
$m_{\phi^0}\ {\lesssim}\ 2m_b$. As will be argued 
from current experimental data in the
next section, we suppose specifically that $m_{\phi^0}$ lies
somewhere between the $\Upsilon(1S)$ and $\Upsilon(2S)$ masses,
i.e.
\begin{equation}
m_{\Upsilon(1S)}\ {\lesssim}\ m_{\phi^0}\ {\lesssim}\ m_{\Upsilon(2S)}
\label{eq:higgsmass}
\end{equation}

Now, we define the mass difference:
${\Delta}m=|m_{\phi^0}-m_{\eta_b}|$, where $\eta_b$
denotes either a $1S$ or a $2S$ state. Accepting for simplicity that
the Higgs boson stands halfway between the mass values of both
resonances, we set ${\Delta}m\ {\simeq}\ 0.25$ GeV
for an order-of-magnitude calculation. 

Hence we write
approximately for the scalar tree-level propagator 
of the $\phi^0$ particle in the process (\ref{eq:channel})
entering in the evaluation of $\Gamma_{\eta_b\to \ell\ell}$,
\begin{equation}
\frac{1}{(m_{\eta_b}^2-m_{\phi^0}^2)^2+m_{\phi^0}^2\Gamma_{\phi^0}^2}\ 
\simeq\ \frac{1}{(m_{\eta_b}^2-m_{\phi^0}^2)^2}\ \simeq\ 
\frac{1}{4\ m_{\eta_b}^2\ ({\Delta}m)^2}
\label{eq:A0prop}
\end{equation}
where the total width of the Higgs boson $\Gamma_{\phi^0}$
has been neglected
assuming that $(\Delta m)^2 \gg \Gamma_{\phi^0}^2$.
We will make a consistency check of this point in subsection 5.1
using numerical values.

Performing a comparison between the widths of 
both leptonic decay processes
(i.e. ${\Gamma}_{\eta_b\to\ell\ell}$ versus
${\Gamma}^{(em)}_{\Upsilon\to\ell\ell}$), one concludes with the aid of 
Eqs.(\ref{eq:A0coupling}-\ref{eq:A0prop}) 
and (\ref{eq:Gamma0}) that 
\begin{equation}
{\Gamma}_{\eta_b\to\ell\ell} \simeq
\frac{3m_b^4m_{\ell}^2(1-4x_{\ell})^{1/2}|R_n(0)|^2\xi_b^2\xi_{\ell}^2}
{2\pi^2(m_{\eta_b}^2-m_{\phi^0}^2)^2v^4} \simeq  
\frac{3\xi_b^2\xi_{\ell}^2}{32\pi^2Q_b^2\alpha^2v^4} \ 
\frac{m_b^4 m_{\ell}^2}{{\Delta}m^2v^4}\ \frac{1}{1+2x_{\ell}} {\times}\ 
{\Gamma}^{(em)}_{\Upsilon\to\ell\ell}
\label{eq:comparison}
\end{equation}

Above we used the non-relativistic 
approximation (more precisely the static approximation) when
assuming null relative momentum of
heavy quarks inside quarkonium, and the same
wave function at the origin for both the $\Upsilon$ vector
state and the $b\bar{b}[{}^1\!S_0^{(1)}]$ intermediate bound state
on account of heavy-quark spin symmetry \cite{bodwin}. Then, the 
decay amplitude squared of the pseudoscalar state
into a  $J^{PC}=0^{++}$ (CP-even) Higgs vanishes and only a $0^{-+}$ 
(CP-odd) Higgs would couple to pseudoscalar quarkonium in this limit. 
Therefore, the Higgs boson hunted in this way
should be properly denoted by $A^0$ and, consequently,
this notation instead of the generic $\phi^0$ 
will be employed for it from now on.

\subsection{Modification of the leptonic BF 
due to a light CP-odd Higgs contribution}

The BF for channel (\ref{eq:channel}) can be readily obtained 
inserting Eq.(\ref{eq:comparison}) into Eq.(\ref{eq:finalwidth2}) 
and afterwards dividing by $\Gamma_{\Upsilon}$, as
\begin{equation}
{\cal B}_{\Upsilon{\rightarrow}\gamma_s\ell\ell}\ \simeq\  
\biggl[\frac{\xi_b^2\xi_{\ell}^2m_b^2m_{\ell}^2\,\Delta E_{hs}^{\,3}}
{8\pi^2{\alpha}(1+2x_{\ell})\Gamma_{\Upsilon}{\Delta}m^2v^4}\biggr] 
\times {\cal B}_{\ell\ell}
\end{equation}
so one can compare
the relative rates by means of the following ratio
\begin{equation}
{\cal R} = \frac{{\cal B}_{\Upsilon{\rightarrow}\gamma_s\ell\ell}}
{{\cal B}_{\ell\ell}} \simeq 
\biggl[\frac{\xi_b^2\xi_{\ell}^2m_b^2\,\Delta E_{hs}^{\,3}}
{8\pi^2{\alpha}(1+2x_{\ell})\Gamma_{\Upsilon}v^4}\biggr] 
\times \frac{m_{\ell}^2}{{\Delta}m^2}
\end{equation}
where we are assuming in the denominator
that the main contribution to the leptonic
channel comes from the photon-exchange graph of Fig.1(a). 
Let us point out once again that, since the $\gamma_s$ 
remains undetected, the NP contribution would be experimentally ascribed
to the leptonic channel of the $\Upsilon$ resonance. Thus, the ratio
(22) represents the fraction of leptonic decays
mediated by a CP-odd Higgs, ultimately
responsible for the breaking of leptonic universality due
to the quadratic mass term $m_{\ell}^2$.

Now, to facilitate the comparison of our results
with other searches for Higgs bosons, we identify in the following
the $\xi_f$ factor with
the 2HDM (Type II) parameter for the universal down-type
fermion coupling to a CP-odd Higgs, i.e. $\xi_b=\xi_{\ell}=\tan{\beta}$
\cite{gunion}. Inserting
numerical values into (22) and keeping the leading term
in $m_{\ell}^2$, one gets the interval
\begin{equation}
{\cal R}\ \simeq\ (1.5{\cdot}10^{-7}-1.2{\cdot}10^{-5})\ {\times}\ 
\tan{}^4\beta\ \times\ m_{\ell}^2
\label{eq:Rvalues}
\end{equation}
where use was made of the approximation
$m_{A^0} \simeq 2m_b \simeq 10$ GeV, and 
the broad range $35-150$ MeV for the possible
hyperfine mass diference $\Delta E_{hs}$ \cite{aleph02};
$m_{\ell}$ is expressed in GeV.

\begin{table} [htb]
\caption{Measured leptonic BF's 
(${\cal B}_{\ell\ell}$) and error bars ($\sigma_{\ell}$) in $\%$, of
$\Upsilon(1S)$ and $\Upsilon(2S)$ (from \cite{pdg}).}
\begin{center}
\begin{tabular}{cccc}
\hline
channel: & $e^+e^-$ & $\mu^+\mu^-$ & $\tau^+\tau^-$  \\
\hline
$\Upsilon(1S)$ & $2.38 \pm 0.11$ & $2.48 \pm 0.06$ & $2.67 \pm 0.16$ \\
\hline
$\Upsilon(2S)$ & $1.34 \pm 0.20$ & $1.31 \pm 0.21$ & $1.7 \pm 1.6$ \\
\hline
\end{tabular}
\end{center}
\end{table}

\subsection{Possible $A^0-\eta_b$ mixing}

Long time ago, the authors of references \cite{haber79,ellis79}
pointed out the possibility of
mixing between a light Higgs (either a CP-even or a CP-odd boson) and 
bottomonium resonances (scalar or pseudoscalar, respectively). Later, 
Drees and Hikasa \cite{drees} made an exhaustive analysis
of the phenomenological consequences of the mixing on the properties
of both resonances and Higgs bosons. 
In view of new and forthcoming data on the bottomonium sector
from B factories \cite{babar,belle}, we are particularly 
interested to apply those ideas 
looking for experimental signatures to provide an additional check
on our conjecture of a light CP-odd Higgs particle.

On the one hand, the mixing can enhance notably
the $gg$ decay mode of the Higgs boson 
\cite{drees}, ultimately increasing 
its total decay width. A net
effect would be an important decrease of the Higgs
tauonic BF (when the $A^0$ mass ranges from 9.4 GeV to 11.0 GeV, especially
if it lies close to the $\eta_b$ masses \cite{drees}).
An exciting experimental consequence 
arises in the search for Higgs particles carried out at LEP: higher 
$\tan{\beta}$ values are allowed than those upper bounds
derived from the analysis 
without considering the mixing \cite{opal,maria1,maria2}.

Interestingly, the mixing with a CP-odd Higgs could also modify the
properties of the pseudoscalar resonances, i.e. the $\eta_{b}$
states. Thus, even for moderate $\tan{\beta}$ one might expect a 
disagreement between forthcoming experimental measurements of
the hyperfine splitting $m_{\Upsilon}-m_{\eta_b}$, 
and theoretical predictions based on potential
models, lattice NRQCD or pQCD 
\cite{aleph02}.
We will come back to this discussion in our numerical
analysis of subsection 5.1.

\section{Lepton universality breaking?}

Let us confront our predictions based on the existence of
a CP-odd Higgs boson with experimental results 
on $\Upsilon$ leptonic decays \cite{pdg} summarized in 
Table 1. Indeed, current data show a slight rise 
of the decay rate with the lepton mass when comparing the
$\tau^+\tau^-$ decay mode with the other two ($e^+e^-$ and
$\mu^+\mu^-$) modes. However, error bars ($\sigma_{\ell}$) are too large 
(especially in the $\Upsilon(2S)$ case) to
permit a thorough check of the lepton mass dependence as
expressed in (\ref{eq:Rvalues}). 
Nevertheless, we have applied a hypothesis test (see appendix A)
to check lepton universality using the $\Upsilon(1S)$ and 
$\Upsilon(2S)$ data displayed in Table 1. 

The null hypothesis (i.e. lepton
universality) is compared against the alternative
hypothesis stemming from the Higgs contribution
predicting a larger (and positive) value of the measured mean of the
${\cal B}_{\ell\ell}$'s differences. Thereby, we conclude 
that lepton universality can be rejected at a $10\%$ level of significance. 
As a cornerstone of this work, 
such slight but measurable variation of the leptonic decay rate   
(by a ${\cal O}(10)\%$ factor) from the electronic/muonic channel to the 
tauonic channel can be interpreted theoretically according to the
2HDM upon a reasonable choice of its parameters (e.g. $\tan{\beta}$)
as we shall see below.

\subsection{Values of $\tan{\beta}$, $A^0-\eta_b$ mixing
and discussion}

In order to
explain the rise of the tauonic BF 
by a $\sim 10\%$ factor w.r.t. the electronic/muonic decay modes, one obtains
from Eq.(\ref{eq:Rvalues}) that 
$\tan{\beta}$ should roughly lie over the range:
\begin{equation}
7\ {\lesssim}\ \tan{\beta}\ {\lesssim}\ 21
\label{eq:tanbeta}
\end{equation}
depending on the value of $\Delta E_{hs}$, namely from 150 MeV 
down to 35 MeV, 
whose limits remain somewhat arbitrary however. The partial
width for the tauonic decay mode of the $\Upsilon(1S)$ 
mediated by the CP-odd Higgs turns
out to be $\Gamma_{\,\Upsilon\to\gamma_s\,\tau\tau} \simeq 140$ eV.

A caveat is in order: the above interval is {\em purely indicative} 
since there are several sources of uncertainty in its calculation,
like the actual mass of the hypothetical
Higgs and not merely the guess made in Eq.(\ref{eq:higgsmass})
or the crude estimate of the probability 
${\cal P}^{\Upsilon}(\eta_b^*\gamma_s)$. In fact, higher values 
of $\tan{\beta}$ were obtained in our earlier work \cite{mas02}
because lower photon energies were
used (i.e. between 10 and 50 MeV). Actually,
letting ${\cal P}^{\Upsilon}(\eta_b^*\gamma_s)$ vary, the
range given in Eq.(\ref{eq:tanbeta}) changes accordingly and
somewhat higher values cannot be ruled out at all. In sum,
our calculations are only approximate and we cannot
claim a well-defined interval for $\tan{\beta}$ but just an
indication on the values needed to interpret a possible
lepton universality breakdown according to our hypothesis 
\footnote{It is also
worthwhile to remark that the range in (\ref{eq:tanbeta}) 
is compatible with the lowest values of $\tan{\beta}$ needed
to interpret the $g-2$ muon anomaly in terms of a
light CP-odd Higgs resulting from a   
two-loop calculation \cite{maria1,maria2,cheung}. 
At present, there is a discrepancy (3.0$\sigma$) between the theoretical
value and the experimental result based on $e^+e^-$ data, {\em but}
only 1.0$\sigma$ when $\tau$ data are used \cite{nyffeler}. Hence the
situation is still unclear to claim for new physics beyond 
the SM from the $g-2$ analysis alone.}. Nevertheless,
we perform below a consistency check of (24)
concerning several partial widths of the $\eta_b$ and $A^0$ particles. 
 
Firstly, let us insert the values of $\tan{\beta}$ 
given by Eq.(\ref{eq:tanbeta}) into 
Eq.(\ref{eq:comparison}) to compute ${\Gamma}_{\eta_b\to\tau\tau}$;
notice that a high value of $\tan{\beta}$  
might yield a large partial width
for the $\eta_b$ resonance, as compared with the
expectation $\Gamma_{\eta_b} \simeq 4$ MeV
obtained in section 3.1. In fact, using the interval given in (24)
one gets ${\Gamma}_{\eta_b\to\tau\tau}$ varying  
from $44$ keV up to 3.56 MeV. Therefore, taking into account the
NP contribution to the total decay rate, 
the Higgs-mediated tauonic BF of the $\eta_b$ resonance
should stay over the range
$ 0.01\ {\lesssim}\ {\Gamma}_{\eta_b\to\tau\tau}/\Gamma_{\eta_b}\ 
{\lesssim}\ 0.5$.

On the other hand, the decay width of a CP-odd Higgs boson into
a tauonic or a $c\bar{c}$ pair in the 2HDM(II)
can be obtained, respectively, from the expressions: \cite{drees}
\begin{eqnarray}
\Gamma(A^0 \rightarrow \tau^+\tau^-) & \simeq & 
\frac{m_{\tau}^2\tan{}^2\beta}{8\pi v^2}\ m_{A^0}\ (1-4x_{\tau})^{1/2} \\
\Gamma(A^0 \rightarrow c\bar{c}) & \simeq & 
\frac{3m_c^2\cot{}^2\beta}{8\pi v^2}\ m_{A^0}\ (1-4x_c)^{1/2}
\end{eqnarray}
where $x_{\tau}=m_{\tau}^2/m_{A^0}^2$ and $x_c=m_c^2/m_{A^0}^2$.
Below open bottom production, even for moderate $\tan{\beta}$,
the $A^0$ decay mode would be dominated by the tauonic channel, i.e. 
\begin{equation}
\Gamma_{A^0}\ \simeq\ \Gamma(A^0\rightarrow \tau^+\tau^-)\ 
\simeq\ 1 - 10\ MeV
\end{equation} 
Thus we can confirm the validity of the approximation
made in Eq.(\ref{eq:A0prop}) for the Higgs propagator, i.e.
$\Delta m^2 \gg \Gamma_{A^0}^2$ (where we tentatively set $\Delta m=250$ MeV).

As commented in subsection 4.2,
there is another interesting consequence of our conjecture related to
bottomonium spectroscopy due to the mixing between a CP-odd Higgs and
$\eta_b$ states. (In appendix B we introduce the notation and
basic formulae.) Indeed, using the values of
$\tan{\beta}$ from Eq.(\ref{eq:tanbeta}) 
the mixing parameter defined in Eq.(B.2)
turns out to be $\delta m^2 \simeq 1.0 - 3.1$ GeV$^2$.
Therefore, such $\eta_b-A^0$ mixing
could induce an observable mass shift of the physical $\eta_b$ states
which would eventually increase the hyperfine splitting
between pseudoscalar and vector resonances w.r.t. a
variety of calculations within the SM \cite{aleph02}.

Let us now write the masses of the mixed 
(physical) states
as a function of the masses of the unmixed states (marked by
a subscript $\lq$0', i.e. $\eta_{b0}$ and $A_0^0$)
with the aid of the expression derived from Eq.(B.5) for
narrow states and $m_{\eta_{b0}} \simeq m_{A_0^0}$,
\begin{equation}
m_{\eta_b,A^0}^2\ \simeq\ \frac{1}{2}(m_{A_0^0}^2
+m_{\eta_{b0}}^2)\ \pm\ \ \frac{1}{2}[(m_{A_0^0}^2
-m_{\eta_{b0}}^2)^2+4(\delta m^2)^2]^{1/2} 
\end{equation}
Taking as a particular case
$\tan{\beta}=16$ ($\delta m^2 \simeq 2.3$ GeV$^2$),  
$m_{\eta_{b0}}=9.4$ GeV and $m_{A_0^0}=9.5$ GeV, 
we get approximately $m_{\eta_b} \simeq 9.32$ GeV and
$m_{A^0} \simeq 9.58$ GeV, 
compatible with our tentative hypothesis on the Higgs mass 
(i.e. ${\Delta}m=|m_{A^0}-m_{\eta_b}| \simeq 0.25$ GeV) and 
the experimental mass
\footnote{The measured mass \cite{aleph02}
of the observed $\eta_b(1S)$ state is $9300 {\pm} 20 {\pm} 20$ MeV, 
indeed slightly smaller than different calculations of
the hyperfine splitting. This measurement 
based on a single event needs confirmation however \cite{cleo03}.} 
so far measured for the  $\eta_b$ meson \cite{aleph02,pdg}. 

\section{Summary}

In this paper we have interpreted a possible breakdown of 
lepton universality in $\Upsilon$ leptonic decays 
(suggested by current experimental data at a 
$10\%$ level of significance) in terms of 
a non-standard CP-odd Higgs boson of mass around $10$ GeV, 
thereby introducing a quadratic dependence on the 
leptonic mass in the corresponding BF's. Higher-order 
corrections within the SM (involving the one-loop
$\eta_b$ decay into two photons, as estimated in appendix C) 
fail by far to explain this effect.

The existence of a CP-odd Higgs of mass about 10 GeV
mixing with pseudoscalar $b\bar{b}$ resonances should display
very clean experimental
signatures and therefore could be easily tested with
present facilities:
\begin{itemize}
\item $\lq\lq$Apparent'' breaking of lepton universality when comparing the
BF of the $\tau^+\tau^-$ decay mode on the one hand, and the 
BF's of the electronic
and muonic modes on the other. Experimental hints of this 
possible signature triggered this work \cite{mas02}.
\item Presence of monoenergetic photons with energy
of order 100 MeV (hence above detection threshold) 
in those events mediated by the CP-odd Higgs boson
(estimated about 10$\%$ of all $\Upsilon$
tauonic decays). This observation would eventually
become a convincing evidence of our working hypothesis. 
\item The $\Upsilon-\eta_b$ hyperfine splitting
larger than SM expectations, caused by the $A^0-\eta_b$ mixing. Also 
a rather large total
width of the $\eta_b$ resonance due to the NP channel (especially 
for the higher values of $\tan{\beta}$ shown in (24)).
\end{itemize}
Although we have focused on a light CP-odd Higgs boson
according to a 2HDM(II) for numerical computations, the main
conclusions may be extended to other pseudoscalar Higgs-like 
particles with analogous phenomenological features, as outlined in the
Introduction. We
thus stress the relevance for checking our conjecture
of new measurements of 
spectroscopy and leptonic decays of the Upsilon family
below $B\bar{B}$ threshold in B factories (BaBar \cite{babar}, 
Belle\cite{belle}) and CLEO \cite{cleo03}.

\subsection*{Acknowledgements} 
I thank G. Bodwin, N. Horwitz, C. Hugonie, 
M. Krawczyk, F. Mart\'{\i}nez-Vidal, 
P. Ruiz and J. Soto for helpful discussions.

\appendix
\appendix\renewcommand{\theequation}{\thesection.\arabic{equation}}
\section*{Appendices}

\section{Lepton universality breaking: hypothesis testing}

\setcounter{equation}{0}

In Table 2 we show the differences between the
branching fractions of the leptonic
channels defined as
${\Delta}_{\ell\ell'}={\cal B}_{\ell'\ell'}-{\cal B}_{\ell\ell}$
($\ell',\ell=e,\mu,\tau$) for both $\Upsilon(1S)$ and $\Upsilon(2S)$
resonances, obtained from Table 1 in the main text. Dividing 
them by their respective experimental errors 
$\sigma_{\ell\ell'}=\sqrt{\sigma_{\ell}^2+\sigma_{\ell'}^2}$ (see also Table 1)
one gets the ratios ${\Delta}_{\ell\ell'}/
\sigma_{\ell\ell'}$. Only four of these quantities
can be considered as independent. Moreover, in view of the
small difference between the electron and the muon masses
as compared with the tau mass, we will base 
our analysis on the comparison between the electron and 
the muon decay modes on one side, versus the tauonic mode on the other side
\footnote{In a prior work \cite{mas02} the
muonic and tauonic modes were confronted with the electronic mode.
Our final conclusion remains the same
as before.}.

In this analysis, we are especially interested 
in the {\em alternative} hypothesis 
based on the existence of a light Higgs boson
enhancing the decay rate as a growing function of
the leptonic squared mass, in opposition to the kinematic factor (2).
Therefore, the {\em region of rejection} for our
statistical test should lie only on {\em one side} (or {\em tail}) 
of the ${\Delta}_{\ell\ell'}/\sigma_{\ell\ell'}$ variable distribution
(i.e. positive values if $m_{\ell'}>m_{\ell}$), in particular
above a  preassigned {\em critical value} \cite{frodesen,martin}.
In other words, we have performed
a {\em one-tailed test} \cite{frodesen,martin} using the sample consisting
of the four independent BF differences between 
the electronic and the muonic channels versus the tauonic
decay mode as explained above (i.e. ${\Delta}_{e\tau}/\sigma_{e\tau}$, 
${\Delta}_{\mu\tau}/\sigma_{\mu\tau}$). For the sake of
simplicity, we
will assume that such differences follow a normal probability distribution
with a mean of  $0.7775$, obtained from 
the four ${\Delta}_{\ell\tau}/{\sigma}_{\ell\tau}$ 
independent values ($\ell=e,\mu$). 

Next, let us define the {\em test statistic}: 
$T={\langle}{\Delta}_{\ell\tau}/{\sigma}_{\ell\tau}{\rangle}{\times}
\sqrt{N} \simeq 1.555$, where $N=4$ stands for the number of 
independent points. Indeed note that we are dealing with a 
Gaussian of unity variance after dividing all differences by
their respective errors.
Now, we will choose the {\em critical value} to be ${\simeq}\ 1.288$
corresponding to a significance level of 10$\%$ in the test.
Lepton universality plays the role  of the {\em null
hypothesis} predicting a mean zero (or slightly less),
against the alternative (composite) hypothesis stemming from
the postulated Higgs contribution predicting a mean value larger than zero
\footnote{We are facing a situation where the null
hypothesis is simple while the alternative is composite but could
be regarded as an aggregate of hypotheses \cite{martin}:
we are assuming normal distributions, with
unit variance and  mean $\mu_0=0$ for
the null hypothesis, while $\mu_a > \mu_0=0$
for the alternative complex hypothesis. A significance level of $10\%$
means that the null hypothesis will be rejected
if the measured mean value of the ${\Delta}_{\ell\ell'}/\sigma_{\ell\ell'}$ 
differences is greater than $\approx 1.288/\sqrt{N}$, 
where $N$ denotes the total number of points. This condition
is equivalent to require that the test statistic $T$ defined
above should be greater than 1.288.}.
Since experimental data imply that
$T>1.288$, we can reject the lepton universality hypothesis
at a $10\%$ level of significance 
\footnote{Let us recall that
the {\em significance level} (or {\em error of the first kind}) represents
the percentage of all decisions such that the null hypothesis
was rejected when it should, in fact, have been accepted 
\cite{frodesen,martin}.}. Certainly, this result alone is
not statistically significant enough to make any serious claim 
about the rejection of the lepton universality hypothesis in
this particular process, but points out the interest to
investigate further the alternative hypothesis stemming from 
our conjecture on the existence of a light Higgs.

\begin{table} [htb]
\caption{All six differences ${\Delta}_{\ell\ell'}$ 
(obtained from Table 1 in the main text) between the leptonic BF's (in $\%$) of
$\Upsilon(1S)$ and $\Upsilon(2S)$ resonances separately.
Subscript ${\ell}\ell'$ denotes the difference between
channels into ${\ell}\bar{\ell}$ and ${\ell}'\bar{\ell}'$ 
lepton pairs
respectively, i.e.
${\Delta}_{\ell\ell'}={\cal B}_{\ell'\ell'}-{\cal B}_{\ell\ell}$; the
${\sigma}_{\ell\ell'}$  values were obtained from Table 1 by summing
the error bars in quadrature, i.e. 
$\sigma_{\ell\ell'}=\sqrt{\sigma_{\ell}^2+\sigma_{\ell'}^2}$. Only two 
${\Delta}_{\ell\ell'}/{\sigma}_{\ell\ell'}$ values for
each resonance can be considered as truly independent, amounting 
altogether to a total number of four independent points.}
\begin{center}
\begin{tabular}{cccc}
\hline
channels & ${\Delta}_{\ell\ell'}$ & $\sigma_{\ell\ell'}$ & 
${\Delta}_{\ell\ell'}/\sigma_{\ell\ell'}$ \\
\hline
$\Upsilon(1S)_{e{\mu}}$  & $0.1$ & $0.125$ & $+0.8$ \\
\hline
$\Upsilon(1S)_{\mu{\tau}}$  & $0.19$ & $0.17$ & $+1.12$ \\
\hline
$\Upsilon(1S)_{e{\tau}}$  & $0.29$ & $0.19$ & $+1.53$ \\
\hline
$\Upsilon(2S)_{e{\mu}}$  & $-0.03$ & $0.29$ & $-0.10$ \\
\hline
$\Upsilon(2S)_{\mu{\tau}}$  & $0.39$ & $1.61$ & $+0.24$ \\
\hline
$\Upsilon(2S)_{e{\tau}}$  & $0.36$ & $1.61$ & $+0.22$ \\
\hline
\end{tabular}
\end{center}
\end{table}

\section{Mixing between a CP-odd Higgs and pseudoscalar resonances 
of bottomonium}
 
\setcounter{equation}{0}

The mixing between Higgs and resonances is
described by the introduction of off-diagonal
elements denoted by $\delta m^2$ in the mass matrix. In our case,
\begin{equation}
{\cal M}_0^2\ =\  
\left(
     \begin{array}{cc}
      m_{A_0^0}^2-im_{A_0^0}\Gamma_{A_0^0} & \delta m^2\\
      \delta m^2 & m_{\eta_{b0}}^2-im_{\eta_{b0}}\Gamma_{\eta_{b0}}      
     \end{array}
\right)
\end{equation}
where the subindex $\lq$0' indicates unmixed states.
The off-diagonal element $\delta m^2$ can be computed
within the framework of a nonrelativistic quark potential model.
For the pseudoscalar case under study, one can write \cite{drees} 
\begin{equation}
\delta m^2\ =\ \xi_b\ 
\biggl(\frac{3m_{\eta_{b}}^3}{4 \pi v^2}\biggr)^{1/2} |R_{\eta_b}(0)|
\end{equation}
Notice that $\delta m^2$ is proportional to $\xi_b$, i.e. $\tan{\beta}$
in the 2HDM(II); high values of the latter implies that mixing
effects can be important over a large mass region.
Substituting numerical values (for the radial wave function at the origin
we used the potential model estimate from \cite{eichten}
$|R_{\eta_b}(0)|^2=$ 6.5 GeV$^3$) one finds (in GeV$^2$ units) 
\begin{equation}
\delta m^2\ \simeq\ 0.146\ \times\ \xi_b
\end{equation}
It is convenient to introduce the complex quantity
\begin{equation}
\Delta^2\ =\ 
\biggl[\frac{1}{4}(m_{A_0^0}^2-m_{\eta_{b0}}^2-im_{A_0^0}\Gamma_{A_0^0}
+im_{\eta_{b0}}\Gamma_{\eta_{b0}})^2+(\delta m^2)^2\biggr]^{1/2}
\end{equation}
and the mixing quantity $\sin{}2\theta = \delta m^2/\Delta^2$, 
where $\theta$ is the (complex) mixing angle of the 
unmixed $\eta_{b0}$ 
resonance and $A_0^0$ Higgs boson giving rise to the
physical eigenstates.
The masses and decay widths of the mixed (physical) states are thus
\begin{equation}
m_{1,2}^2-im_{1,2}\Gamma_{1,2}\ =\ \frac{1}{2}
(m_{A_0^0}^2+m_{\eta_{b0}}^2-im_{\eta_{b0}}
\Gamma_{\eta_{b0}}-im_{A_0^0}\Gamma_{A_0^0}
\ {\pm}\ \Delta^2)
\end{equation}
where subscripts $1,2$ refer to a Higgs-like state and a
resonance state respectively, if $m_{A_0^0}>m_{\eta_{b0}}$; the converse
if  $m_{A_0^0}<m_{\eta_{b0}}$.

\begin{figure}[htb]
\centerline{\hbox{
 \psfig{file=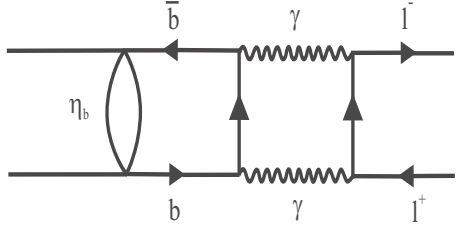,height=3.cm,width=6cm} }}
\caption{One-loop process within the SM potentially contributing to the
dependence of the leptonic BF on the mass of the final-state leptons.}
\end{figure}

\section{$\eta_b \rightarrow \ell^+\ell^-$: one-loop calculation 
within the Standard Model}

\setcounter{equation}{0}

In this appendix we consider within the SM an alternative possibility
to the Higgs conjecture of a rising 
leptonic BF of bottomonium with the leptonic mass, based on the 
electromagnetic
decay into a lepton pair of the $\eta_b$ state subsequent to the
magnetic dipole transition advocated in this work 
(see Fig.2). 
Since the width for the $\eta_b$ 
decay into two photons has been recently
calculated  elsewhere \cite{fabiano}, obtaining
the range $\Gamma(\eta_b \rightarrow \gamma\gamma) \simeq 0.4 - 0.5$ 
keV, we can use the following estimate according to the SM:
\begin{equation}
\Gamma_{SM}(\eta_b \rightarrow \ell^+\ell^-) \simeq 
 \Gamma(\eta_b\ \rightarrow\ \gamma\gamma) {\times} 
\frac{\alpha^2m_{\ell}^2}{m_{\eta_b}^2} \frac{1}{2\lambda}\ 
\biggl(\log{\frac{1+\lambda}{1-\lambda}}\biggr)^2 \ll 
{\Gamma}_{\eta_b\to\ell\ell}
\end{equation}
where $\lambda=(1-4m_{\ell}^2/m_{\eta_b}^2)^{1/2}$. The above equation
corresponds to the unitary bound due to the absorptive contribution
of the two-photon exchange. The last inequality is readily
obtained by setting the numerical values for $\tan{\beta}$
used in this work to get $\Gamma_{\eta_b\to\ell\ell}$ for
any leptonic species.

\end{document}